\documentstyle[12pt]{article}
\def\beq{\begin{equation}}   \def\eeq{\end{equation}}
 \newcommand{\gsim}{\lower.7ex\hbox{$
\;\stackrel{\textstyle>}{\sim}\;$}}
\newcommand{\lsim}{\lower.7ex\hbox{$
\;\stackrel{\textstyle<}{\sim}\;$}}

\newcommand{\ra}{\rightarrow}

\newcommand{\La}{\overline{\Lambda}}
\newcommand{\Lam}{\Lambda_{\rm QCD}}

\newcommand{\re}[1]{Ref.~\cite{#1}}
\newcommand{\eq}[1]{Eq.\hspace*{.1em}(\ref{#1})}

\newcommand{\de}{\delta}

\newcommand{\as}{\alpha_s}

\newcommand{\matel}[3]{\langle #1|#2|#3\rangle}

 \newcommand{\GeV}{\,\mbox{GeV}}

\begin{document}

\def\lsim{\mathrel{\rlap{\lower3pt\hbox{\hskip0pt$\sim$}}
    \raise1pt\hbox{$<$}}}         
\def\gsim{\mathrel{\rlap{\lower4pt\hbox{\hskip1pt$\sim$}}
    \raise1pt\hbox{$>$}}}         

\begin{titlepage}
\renewcommand{\thefootnote}{\fnsymbol{footnote}}

\begin{flushright}
CERN-TH/96-191\\
TPI-MINN-96/13-T\\
UMN-TH-1506-96\\
UND-HEP-97-BIG\hspace*{.1em}02\\
hep-ph/9704245\\
\end{flushright}
\vspace{.3cm}

\begin{center} 
{\Large
{\bf High Power $n$ of $m_b$ in Beauty Widths and 
$n=5 \rightarrow \infty\;$ Limit}}
\end{center}
\vspace*{.3cm}
\begin{center} {\Large
I. Bigi$^{\:a}$, M. Shifman$^{\:b}$, N. Uraltsev$^{\:a{\rm -}d}$ and 
A. Vainshtein$^{\:b,e}$} \\
\vspace{.4cm}
{\normalsize 
$^a${\it Dept.of Physics,
Univ. of Notre Dame du
Lac, Notre Dame, IN 46556, U.S.A.\\
$^b$ {\it  Theoretical Physics Institute, Univ. of Minnesota,
Minneapolis, MN 55455}\\
$^c$ {\it TH Division, CERN, CH 1211 Geneva 23, Switzerland}\\
$^d$ {\it Petersburg Nuclear Physics Institute,
Gatchina, St.~Petersburg 188350, Russia\footnote{Permanent 
address}
}}\\
$^e$ {\it Budker Institute of Nuclear Physics, Novosibirsk
630090, Russia}\vspace*{.8cm}\\
}

{\Large{\bf Abstract}}
\end{center}
\vspace*{.2cm}

\noindent
The leading term in the semileptonic width of heavy flavor  hadrons
depends on the fifth power of the heavy quark mass. We  present
an analysis where this power can  be self-consistently treated as a
free parameter $n$ and the width can be studied in
the limit   $n\ra \infty$.
The resulting expansion elucidates why the small  velocity (SV)
treatment  is relevant for the inclusive semileptonic $b\ra c$
transition. The extended SV  limit (ESV limit) is introduced. The
 leading terms in the perturbative $\alpha_s$ expansion enhanced 
by
powers of $n$ are automatically resummed by using the low-scale
Euclidean mass.  The large-$n$ treatment explains why the scales
of order
 $m_b/n$ are appropriate. 
On the other hand, the scale cannot be too small since
the factorially divergent
perturbative corrections 
associated with running of $\alpha_s$ 
show up. 
Both requirements are met if we use the  short-distance
mass normalized at a scale around $m_b/n \sim 1\GeV$. A 
convenient
definition of such low-scale OPE-compatible masses is briefly 
discussed.

\vfill
\noindent
CERN-TH/96-191\\
July 1996/March 1997
\end{titlepage}
\addtocounter{footnote}{-1}

\newpage

\section{Introduction}

A true measure of our understanding of heavy flavor 
decays is provided by our ability to extract accurate  
values for fundamental quantities, like the 
CKM parameters, from the data. The inclusive semileptonic 
widths of $B$ mesons depend directly on $|V_{cb}|$ and $|V_{ub}|$,
with the added bonus that they are amenable to the OPE treatment
\cite{svold,cgg,mirage}, and 
the non-perturbative corrections to 
$\Gamma_{\rm sl}(B)$ are of order $1/m_b^2$ \cite{buv,bs,dpf,prl}, 
i.e. 
small. Moreover, they can be expressed, in a model independent 
way, in terms of fundamental parameters of the heavy quark theory 
-- $\mu_\pi^2$ and $\mu_G^2$ -- which are known  with  relatively
small uncertainties. 

Therefore, the main problem in the program of a precise 
determination of, say, $|V_{cb}|$ from $\Gamma _{\rm sl}(B)$
is the theoretical understanding of the {\em perturbative} QCD 
corrections and
the heavy quark masses.
The  theoretical 
expression for $\Gamma _{\rm sl}(B)$ depends  on a high power of 
the quark 
masses $m_b$ (and $m_c$):  
$\Gamma_{\rm sl}\propto m_b^n$ with $n=5$. 
The quark masses are not observables, and the uncertainties in their 
values are 
magnified by the power $n$. 
For this reason it is often  thought that the
perturbative corrections in the inclusive widths may go beyond 
theoretical
control.

The quark masses in the field theory are not constant but rather 
depend on the
scale and, in this respect, are similar to other couplings like $\as$ 
which
define the theory. Of course, $m_Q$
and $\as$ enter the expansion differently. Moreover, in contrast to 
$\as$ in QCD the heavy quark mass
$m_Q$ has a finite infrared limit to any order in the
perturbation theory, $m_Q^{\rm pole}$, which is routinely used in 
the
calculations.

If $\Gamma _{\rm sl}(B)$ is expressed in terms of the pole mass of 
the
$b$ quark treated as a given number, the expression for $\Gamma 
_{\rm
sl}(B)$ contains a factorially  divergent series  in powers of
$\alpha_s$ 
$$
\Gamma _{\rm sl} \sim \sum_k k!\left(\frac{\beta_0}{2} \frac{\alpha_s}{\pi}
\right)^k
$$ 
due  to the $1/m_Q$ infrared (IR)
renormalon \cite{pole,bbz,bbbsl}. This series gives rise to an 
unavoidable
uncertainty which is linear  rather than quadratic  in $1/m_b$. The
related perturbative corrections are significant already in  low 
orders.
They are directly associated with running of $\as$ and their growth
reflects the fact that the strong coupling evolves into the
nonperturbative domain at a low enough scale.  A strategy allowing 
one
to circumvent  this potentially large  uncertainty was indicated in 
Refs.~\cite{pole,bbz} -- instead of the pole quark mass one should 
pass to a
Euclidean mass, according to Wilson's OPE. This mass is peeled off 
from
the IR part; the IR domain is also absent from the width up to effects
$1/m_Q^2$, and the (IR-related) factorial divergence disappears. 
Following this observation, it became routine to express $\Gamma 
_{\rm
sl}(B)$ in terms of the Euclidean quark masses in the $\overline{\rm
MS}$ scheme, say, $\overline{m}_b (m_b)$. 

Killing IR renormalons on this route, however, one does not 
necessary
have a fast convergent perturbative series: in general, it contains
corrections of the type $(n\as)^k$ which are not related to running of
$\as$ -- they are present even for the vanishing $\beta$ function. 
We will see that such large corrections inevitably appear if one 
works with the $\overline{\rm
MS}$ masses like $\overline{m}_b (m_b)$ and $\overline{m}_c 
(m_c)$. Due
to the high value of the power $n$ they constitutes quite an obvious
menace to the precision of the theoretical predictions. The standard
alternative procedure of treating inclusive heavy quark decays, 
which 
gradually evolved in the quest for higher accuracy, is thus plagued 
by
its own problems.

The aim of the present work is to turn   vices  into virtues, by
treating $n$ as a free  parameter and developing a  $1/n$ expansion. 
In
this respect our approach is conceptually similar to  the $1/N_c$  
expansion 
or the expansion in the dimensions of the space-time, etc. which are 
quite
common in various applications of field  theory (for a discussion of 
the
virtues of  expansion in `artificial' parameters,  see \cite{witten}).
The main advantage is  the emergence of  a qualitative picture which
guides the theoretical estimates in the absence of much more
sophisticated explicit higher-order calculations. 

Certain unnaturalness of the $\overline{\rm MS}$ masses 
normalized,
say, at $\mu=m_b$, for inclusive widths is rather obvious before a 
dedicated analysis, in
particular in $b\ra c\, \ell \nu$ inclusive decays. The 
maximal energy fed into
the final hadronic system and available for exciting the final 
hadronic
states, which determines the ``hardness" of the
process, is limited by $m_b-m_c\simeq 3.5 \GeV$. Moreover, in a
typical decay event, leptons carry away a significant energy 
$E_{\ell\nu}>\sqrt{q^2}$, since the lepton phase space emphasizes
the larger
$q^2$. This is illustrated in Fig.~1 where the distribution over
invariant mass of the lepton pair $\sqrt{q^2}$ and energy release
$E_{\rm r} = m_b-m_c-\sqrt{q^2}$ is shown. The situation is less 
obvious
{\em a priori} for $b\ra u$, but again the typical energy of the final
state hadrons is manifestly smaller than $m_b$. We use the 
parameter
$n$ to quantify these intuitive observations.

\begin{figure}
\vspace{4.8cm}
\includegraphics{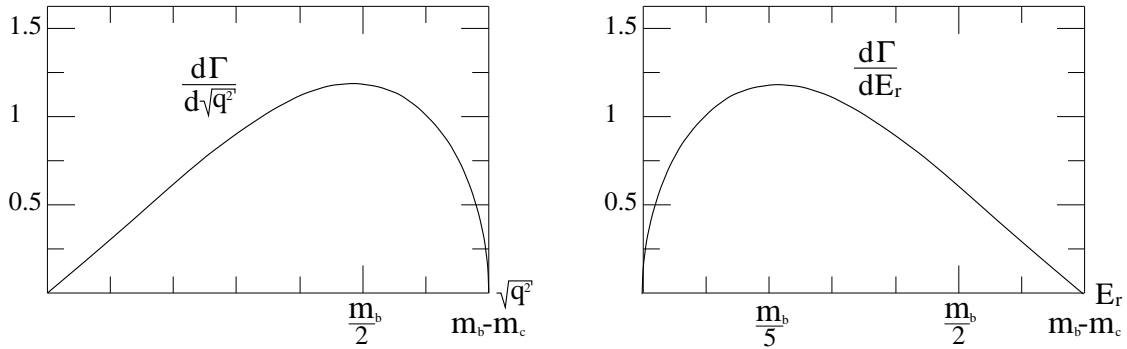}
\caption{
The decay distribution over invariant mass of the lepton pair and 
over 
energy release $E_{\rm r}=m_b-m_c-\left(q^2\right)^{1/2}$ in $b\ra 
c\,\ell\nu$
at $m_c/m_b=0.3$.}
\end{figure}

Since the power $n=5$ is of a purely kinematic origin, technically it
is quite easy to make $n$ a free parameter. To this end it is
sufficient, for instance, to modify the lepton current by introducing
fictitious additional leptons emitted by the $W$ boson, along with the
standard $\ell\nu$ pair. At the very  end the number of these
fictitious leptons is to be put to zero.

The emergence of the large perturbative terms containing powers of 
$n$
in a general calculation is rather obvious and can be illustrated in
the following way. Consider, for example, the $b\ra u$ decay rate,  
and
use the pole mass $m_b^{\rm pole}$. The series will have a factorial 
divergence
in high orders due to the $1/m$ IR renormalon, but we are not 
concerned about
this fact now since we reside in purely perturbative domain and we 
do not study high
orders $k\sim 1/\as$ of the perturbation theory. We can consider, 
for example, 
the case of vanishing $\beta$ function when nothing prevents one 
from 
defining the pole mass with the arbitrary precision.

Let us assume then that using the pole mass, $m_b^{\rm pole}$, the
perturbative expansion of the width 
\beq
\Gamma\;\simeq\; d_n\, m_b^n \, A^{\rm pt}(\as)\;=\;
 d_n\, m_b^n 
\,\left[1+a_1\frac{\as}{\pi}+a_2\left(\frac{\as}{\pi}\right)^2
+\,...\;\right]
\label{i1}
\eeq
has all coefficients $a_k$ completely $n$-independent 
(we will show below that
it is the case in the leading-$n$ approximation).
Then, being  expressed in terms of a different mass $\tilde m_b$,
\beq
\tilde m_b \;=\; m_b \left(1-c \frac{\as}{\pi}\right)
\label{i2}
\eeq
one has 
\beq
\Gamma\;\simeq\; d_n\, \tilde m_b^n \, \tilde A^{\rm pt}(\as)
\label{i3}
\eeq
with
$$
\tilde A^{\rm pt}(\as)\;=\;\frac{A^{\rm pt}(\as)}{\left(1-c
\frac{\as}{\pi}\right)^n}\;=
$$
\beq
1+(n c +a_1)\frac{\as}{\pi}+
\left(\frac{n(n+1)}{2}c^2+n\, c a_1+a_2\right)
\left(\frac{\as}{\pi}\right)^2 +\,...\;
\;\;.
\label{i4}
\eeq
Just using a different mass generates $n$-enhanced terms $\sim 
(n\as)^k$. In
the case of the $\overline{\rm MS}$ mass one has, basically, $c=4/3$. 
In order
to have a good control over the perturbative corrections in the actual 
width,
one needs to  resum these terms (at least, partially). This can be 
readily done.

These $n$-enhanced perturbative corrections are only one, purely 
perturbative,
aspect of a general large-$n$ picture. The total width is determined 
by
integrating $n$-independent hadronic structure functions with the 
$n$-dependent
kinematic factors, the latter being saturated at large $\sqrt{q^2}$ 
close 
to the
energy release $m_b-m_c$. As a result, the energy scale defining the 
effective 
width of integration, is essentially smaller than the energy release; a
new, lower momentum scale automatically emerges in the problem 
at large $n$.

The question of the characteristic scale in inclusive 
decays was
discussed  in the literature. The discussion   was focused, however, 
almost exclusively on
the problem of choosing the normalization point for the running 
coupling $\as$.
We emphasize that for the inclusive decays it is a secondary 
question; the
principal one is the normalization point for quark masses which run 
as well.
Although their relative variation is smaller, this effect is enhanced, in
particular, by the fifth power occurring in the width.

With $n=5$ treated as a free parameter, 
we arrive at the following results concerning semileptonic 
$B$ decays driven by $b \ra l \nu q$, $q=c$ or $u$:
  
\begin{enumerate}  
\item The leading $n$-enhanced perturbative corrections to
the total width are readily resummed by using the low-scale quark 
masses. No
significant uncertainty in the perturbative corrections is left in the 
widths.
\item   The surprising proximity of the $b\ra c\ell \nu $  transition 
to  the
small velocity (SV)  limit \cite{SV} which seems to hold in spite of 
the fact
that $m_c^2/m_b^2\ll 1$, becomes understood. The typical  
kinematics of the
transition are governed by the parameter  $(m_b - m_c)/nm_c$ 
rather than by 
$(m_b - m_c)/m_c$. This ``extended SV" parameter shows  why the 
SV expansion is
relevant for actual $b \ra c$  decays. 
\item The `large-$n$' remarks are
relevant even if $\as$ does not run. In actual QCD, the better control 
of
the perturbative effects meets a conflict of interests: the 
normalization point
for masses cannot be taken either very low or too high, thus neither
$\overline{\rm MS}$ nor pole masses are suitable. We discuss a 
proper way to
define a short-distance heavy quark mass $m_Q(\mu)$ with $\mu \ll 
m_Q$ which is
a must for nonrelativistic expansion in QCD. A similar definition of 
the
effective higher-dimension operators is also given. 
\item  The limits $m_Q \ra
\infty$ and $n\ra \infty$  cannot freely be interchanged. Quite 
different
physical situations  arise in the two cases, as indicated later. The 
perturbative regime takes place if $(m_b-m_c)/n \gg \Lam$, and the 
process is not truly
short-distance if the opposite holds. Nevertheless, at $m_c\gg \Lam$ 
even in
the latter limit  the width is determined perturbatively up to effects 
$\sim
\Lam^2/m_c^2$. \end{enumerate} 

We hasten to add that analyzing only terms leading in $n$  may not 
be
fully adequate for sufficiently accurate predictions in the real 
world where $n=5$, which turns out to be not large enough. Still, it
is advantageous to start from the large $n$ limit and subsequently
include (some of  the) subleading terms.

\section{The theoretical framework: large-$n$ expansion in inclusive 
semileptonic decays}

The new tool we bring  to bear here is the following:   the actual
value $n=5$ of the power $n$ in the width arises largely through 
the integration over the phase space for the lepton pair. One  can
draw up simple physical scenarios where   $n$ appears as a free
parameter and the limit $n\ra \infty$  can be analyzed. The simplest
example is provided by  $l$ (massless) scalar ``leptons" $\phi$,
emitted in the weak vertex of the lepton-hadron interaction,
\beq
{\cal L}_{\rm weak} = \frac{G_l}{\sqrt{2}} V_{Qq}\phi^{\, l}
\bar\ell\gamma_\alpha (1-\gamma_5)\nu_\ell\,\,
\bar{q}\gamma^\alpha (1-\gamma_5)Q\, ,
\label{Lweak}
\eeq
where $G_l$ generalizes the Fermi coupling constant to $l\neq 0$.
At $l = 0$ the coupling $G_0 = G_F$. Then by
dimensional counting one  concludes that  $\Gamma (Q) \sim
|G_l|^2\cdot m_Q^n$, where $ n =2l + 5 $. It should be kept  in mind
that the details of the physical  scenario are not essential for our
conclusions.   

We will discuss $b\ra q$ transitions with an arbitrary mass ratio,
$m_q/m_b$, thus incorporating both  $b\ra u$ and $b\ra c$ decays. 
Following Refs.~\cite{cgg,KOYRAKH} we introduce a  hadronic tensor
$h_{\mu\nu}(q_0, q^2)$, its absorptive part 
$W_{\mu\nu}(q_0,q^2)=\frac{1}{i}\,{\rm disc}\, 
h_{\mu\nu}(q_0,q^2)$
and its decomposition into five covariants  with structure functions
$w_i(q_0, q^2)$, $i=1,...,5$. Variables $q_0$ and $q^2$ are the energy
and the effective mass of lepton pair. Only $w_1$ and  $w_2$ 
contribute
when lepton masses are neglected and the semileptonic decay width
is then given by 
$$
\Gamma_{\rm sl}\equiv |V_{cb}|^2 \frac{G_F^2}{8\pi^3} 
\cdot \gamma \;,
$$
\beq 
\gamma 
=\frac{1}{2\pi}
\int_0^{q^2_{\rm max}}\,{\rm d} q^2\int_{\sqrt{q^2}}^{q_{0\, \rm 
max}}\, 
{\rm d}q_0\,  \sqrt{q_0^2-q^2} 
\left\{q^2w_1(q_0,q^2)+\frac{1}{3}(q_0^2-q^2)w_2(q_0,q^2)\right\}\; 
, 
\label{1}
\eeq
where
$$
q^2_{\rm max} = (M_B-M_D)^2\, , \,
\,\, q_{0\, \rm max} = \frac{M_B^2+q^2-M_D^2}{2M_B}\, .
$$
Note that the upper limits of integration over $q^2$ and $q_0$ 
are determined by vanishing of the structure functions for
the invariant mass of the hadronic system less than $M_D^2$. 
The lower limit of integration over  $q_0$ and $q^2$  
 is due to the constraints on the momentum of the
lepton pair. It has nothing to do with the properties of
the structure functions. In particular, the structure functions
exist also in the scattering channel where the constraint
on the lepton momentum would be different. In what follows we
will see that the distinction between the
leptonic and hadronic kinematical constraints
is important.

By adding $l$ extra scalar ``leptons" emitted in the weak vertex
we make $n=5+2l$  a free parameter.  The quantity $\gamma$ 
 then becomes $n$ dependent, $\gamma \ra 
\gamma(n)$,  
$$ 
\gamma(n)=
\frac{1}{2\pi}
\int_0^{q^2_{\rm max}}\,{\rm d}q^2\, (q^2)^{l}
\int_{\sqrt{q^2}}^{q_{0\, \rm max}}\, {\rm d}q_0\:
 \sqrt{q_0^2-q^2} 
\left\{q^2w_1(q_0,q^2)+\right.
$$
\beq
\left. \frac{1}{3}( q_0^2 - q^2)
w_2(q_0,q^2)\right\}\, .
\label{2}
\eeq
In the real world $l=0$, $n=5$,  but as 
far as the QCD part is concerned
we are free to consider any value of $l$. The extra factor $(q^2)^l$
appeared due to the phase space of the scalar ``leptons".\footnote{
Strictly speaking, the  extra ``leptons'' produce  a change in 
the leptonic tensor, in particular,
making it non-transversal. This variation leads, in turn, to a
different relative weight of the structure function $w_2$, and the 
emergence of the structure functions $w_4$ and $w_5$ in
the total probability.  These complications
are irrelevant, and we omit them. We also omit an overall $l$ 
dependent
numerical factor, which merely redefines $G_l$.}

A different choice of variables is more convenient for our purposes.
Instead of $q_0$ and $q^2$ we will use
$\epsilon$ and $T$, 
\beq
\epsilon =M_B-q_0 -\sqrt{M_D^2+q_0^2 -q^2}\; ,\;\;\;
T=\sqrt{M_D^2+q_0^2 - q^2}-M_D\;.
\label{opred}
\eeq
The variable $T$ is simply related to the spatial momentum $\vec 
q$.
More exactly, $T$ has the meaning of the
minimal kinetic energy of the hadronic system for the
given value of $\vec q$.  This minimum is achieved
when the $D$ meson is produced.
The variable $\epsilon$ has the meaning of the excitation
energy, $\epsilon = (M_X-M_D)+(T_X-T)$ where $T_X
=\sqrt{M_X^2+q_0^2 - q^2}-M_X$
is the kinetic energy of the excited state with mass
$M_X$.
The following notations are consistently used below:
$$
\Delta=M_B-M_D\; , \;\;\;
\vec q\,^2=q_0^2-q^2\; , \;\;\; |\vec q\,|\equiv \sqrt{\vec q\,^2}\; .
$$ 
In terms of  the new variables one obtains
$$
\gamma(n) 
=\frac{1}{\pi}
\int_0^{T_{\rm max}}  {\rm d}T\,(T+M_D)\sqrt{T^2 +2M_D T}
\int_{0}^{\epsilon_{\rm max}} {{\rm d}\epsilon}
\left( \Delta^2-2M_B T-
2\Delta\epsilon+2T\epsilon+\epsilon^2\right)^l
$$
\beq
\left\{ (\Delta^2-2M_B T-2\Delta\epsilon+2T\epsilon+\epsilon^2) 
 w_1 + 
\frac{1}{3}(T^2 +2M_D T) w_2 \right\}
\label{3}
\eeq
where 
$$
T_{\rm max}= \frac{\Delta^2}{2 M_B}\,, \;\; \;
\epsilon_{\rm max}= \Delta - T -\sqrt{T^2 + 2M_D T}\;\, ;
$$
it is implied that the structure functions $w_{1,2}$ depend on 
$\epsilon$ and $T$.

\subsection{Cancelation of the infrared contribution}

Before submerging  in the large $n$ limit we will address a question
which naturally comes to one's mind immediately upon inspection of 
Eq.~(\ref{1}) or Eq.~(\ref{2}). Indeed, these expressions give the total
decay probabilities in terms of an  integral over the physical spectral
densities over the physical phase space. Both factors depend on the
meson masses, and know nothing about the quark mass. This is 
especially
clear in the case of the phase space, which seems to carry the 
main
dependence for large $n$. And yet, in the total probability the 
dependence
on the meson masses must disappear, and the heavy quark mass 
must emerge
as a relevant parameter. In other words, the large-distance
contributions responsible for making the meson mass out of the 
quark one
must cancel each other. 

It is instructive to trace how this
cancelation occurs. We will study the issue for
arbitrary value of $l$, assuming for simplicity that
the final quark mass vanishes (i.e. $b\ra u$ transition).
This assumption is not crucial; one can consider
arbitrary ratio $m_c/m_b$, with  similar conclusions.
Since the main purpose of this section is methodical
we will limit our analysis to terms linear in
$\La$. The basic theoretical tool allowing one to trace
how $M_B^n$ in the decay probability is substituted by
$m_b^n$, as a result of cancelation of the infrared
contributions residing in $M_B$
and in the integrals over the spectral densities,
 is the heavy flavor sum rules
presented in great detail in Ref. \cite{optical}, 
see Eqs.~(130) -- (135).

We start from expanding the integrand in Eq.
(\ref{3}) in $\epsilon$, keeping the terms of the zeroth
and first order in $\epsilon$,
$$
\gamma(n) 
=\frac{1}{\pi}
\int_0^{M_B/2} {\rm d} T\,T^2 M_B^l( M_B -2 T)^l\times
$$
$$
\left[ M_B( M_B-2 T)
\int_{0}^{M_B -2T}{{\rm d}\epsilon} \,  w_1 
-2(l+1)(M_B -T) \int_{0}^{M_B -2T} {{\rm d}\epsilon}   \, \epsilon\, 
w_1
\right] +
$$
$$
\frac{1}{3 \pi} 
 \int_0^{M_B/2}  {\rm d}T\,T^4 M_B^{l-1}( M_B -2 T)^{l-1}\times
$$
\beq
\left[ M_B( M_B-2 T)
\int_{0}^{M_B -2T}{{\rm d}\epsilon} \,  w_2 
-2l(M_B -T) \int_{0}^{M_B -2T}  {{\rm d}\epsilon} \, \epsilon\, w_2
\right] \, .
\label{new3}
\eeq
The integrals over $\epsilon$ are given by the sum rules
mentioned above,
\beq
\frac{1}{2\pi } \int {{\rm d}\epsilon}\,  w_1 =1\, ,\,\,\,
\frac{1}{2\pi } \int {{\rm d}\epsilon}\, \epsilon\, w_1 =
M_B-m_b =\La\, ,
\label{sr1}
\eeq
and 
\beq
\frac{1}{2\pi } \int {{\rm d}\epsilon}\,  w_2 =\frac{2m_b}{T}\, ,\,\,\,
\frac{1}{2\pi } \int {{\rm d}\epsilon}\, \epsilon\, w_2 =
\frac{2m_b}{T} \La\, .
\label{sr2}
\eeq
The expressions (\ref{sr1}) and (\ref{sr2})
imply that the integration over $\epsilon$ is saturated at
small $\epsilon$, of order $\La$. 
After substituting the sum rules in Eq. (\ref{new3})
and integrating over $T$ we arrive at
$$
\gamma (n) = \frac{M_B^{2l+5}}{2(l+4)(l+3)(l+2)}
\left[ 1 -(2l+5)\frac{\La}{M_B}\right] +
$$
\beq
\frac{m_b M_B^{2l+4}}{2(l+4)(l+3)(l+2)(l+1)}
\left[ 1 -(2l+4)\frac{\La}{M_B}\right]\, .
\label{sr3}
\eeq
The first term here comes from $w_1$ and the second from $w_2$.
This expression explicitly demonstrates that
the meson mass is substituted by the quark one
at the level of $1/m$ terms under consideration,
$M_B (1-\La /M_B ) = m_b$. 

Thus, the cancelation of the infrared contribution is evident.
It is worth emphasizing that the infrared contributions
we speak of need not necessarily be of
nonperturbative nature. The cancelation of infrared
$1/m_b$ effects takes place for perturbative contributions as well
as long as they correspond to sufficiently small momenta,   
$ \lsim m_b/n$.

Let us parenthetically note that Eq. (\ref{sr3})
is valid for any $l$; in particular, we can put $l$ equal to zero.
One may wonder how one can get in this case a non-vanishing 
correction
associated with $\int {\rm d}\epsilon\, \epsilon \, w_2$, while the
weight factor for  $w_2$ in the original integrand (\ref{3}) at $l=0$ 
seemingly has no dependence on $\epsilon$ at all. 
This case is
actually singular: the domain of $T\ra M_B/2$ contributes to the 
integral at the
level $\Lam/M_B$. For such $T$ the interval of integration 
over $\epsilon$ shrinks to zero, but nevertheless the integral of 
$w_2$ stays 
finite (unity) according to Eq.~(\ref{sr2}); this signals a
singularity. It is directly seen from Eq.~(\ref{new3}) where at $l=0$
one has logarithmic divergence at $T\ra M_B/2$. 
The most simple correct way to deal with  $w_2$ at $l=0$
is inserting a step-function $\theta (M_B -2T - \epsilon)$ in the integrand
instead of the upper limits of integration. This
step-function does provide an $\epsilon$ dependence,
the term linear in $\epsilon$ is $ \epsilon \, \delta (M_B -2T)$.
At non-vanishing values of $l$ the $\delta$-function above produces no 
effect. As usual, analytic continuation from the nonsingular case 
$l>0$ to $l=0$ leads to the same result.  

In principle, it is instructive to trace the
$l$ dependence of other non-perturbative terms,
e.g. those proportional to $\mu_\pi^2$ and $\mu_G^2$ which are 
known for
$b\ra u$ at arbitrary $l$ from explicit calculations of
Refs.~\cite{buv,bs,dpf} and have a very simple form. We will not 
dwell on 
this issue here.

Our prime concern in this work is 
the interplay of the
infrared effects in the  perturbative corrections. The main
new element appearing in the analysis of the
perturbative corrections is the fact that
the integrals over $\epsilon$ are not saturated
in the domain $\epsilon \sim \La$.
That is why we need to introduce the
normalization point $\mu$
which will divide the entire range of the
$\epsilon$ integration in two domains,
$\epsilon > \mu$ and $\epsilon < \mu$.

The domain $\epsilon < \mu$ provides us
with a clear-cut physical definition
of $\La$, which becomes $\mu$ dependent, 
\beq
\La (\mu ) +
\frac{\mu_G^2}{2|\vec q\, |} +\frac{\mu_\pi^2-\mu_G^2}{3|\vec q\, |}
\left( 1-\frac{|\vec q \,|}{m_b}\right) +
{\cal O}\left( \frac{\Lambda_{\rm QCD}^3}{\vec q^{\,2}}
\right)  
=\frac{\int_0^\mu {{\rm d}\epsilon}\, \epsilon\, w_1^{b\ra u}
(\epsilon , |\vec q \,|)}{\int_0^\mu {{\rm d}\epsilon}\, w_1^{b\ra u}
(\epsilon , |\vec q \,|)}\, ,
\label{DI}
\eeq
where the variable $\epsilon$ is defined by Eq.~(\ref{opred})
with $M_D$ set equal to zero, the second variable 
$T=|\vec q \,|$ in the $b\ra u$
transition. Equation (\ref{DI}) is a consequence of the
sum rules derived in Ref. \cite{optical}. There, the integrals over the 
spectral densities are given in the form of a ``condensate"  expansion.
The currents inducing given spectral densities (and the spectral 
densities themselves) acquire certain normalization factors due to  
short-distance renormalization  of the weak vertex, which are 
irrelevant for our present purposes.
To get rid of these normalization factors we consider the ratio
of the sum rules. The power corrections to the denominator
 in the right-hand side of Eq. (\ref{DI}) shows up only in 
${\cal O}\left( {\Lambda_{\rm QCD}^3}/{\vec q^{\,2}}
\right) $ terms.  

The quark mass, $m_b (\mu ) = M_B - \La (\mu ) $,
becomes in this way a well-defined and experimentally measurable
quantity. 
A very similar definition of $\La (\mu) $ does exist also
in the $b\ra c$ transition,
through the integral over $w_1^{b\ra c}$,
 see Ref. \cite{optical}. This definition generalizes
the Voloshin sum rule \cite{Volopt} to the arbitrary values of $\vec 
q $,
and includes  corrections ${\cal O}(\Lambda_{\rm QCD}^2/m)$. 
See Section 4 for the further discussion of $\La (\mu) $ definition.

\subsection{The large $n$ limit}

Now when we understand how the infrared parts cancel,
we are ready to address the practical issue of
what particular normalization point $\mu$
is convenient to use in the analysis of the
total widths. Certainly, the theoretical predictions
can be given for any value of $\mu \gg \Lambda_{\rm QCD}$.
Our goal is to choose $\mu$ in such a way
that the domain of momenta above $\mu$
does not give the enhanced perturbative corrections. 
Then all  perturbative contributions enhanced by large $n$
will be automatically included in the definition of 
$m_b (\mu )$.
 
 To this end we invoke the large $n$ limit,
as was explained in the Introduction.
For $l \gg 1$ one has 
$$
(\Delta^2-2M_B T-2\Delta\epsilon+2T\epsilon+\epsilon^2)^l\simeq
\Delta^{2l}\:\exp{\left[-2l\left(\frac{TM_B}{\Delta^2}+
\frac{\epsilon}{\Delta}-\frac{T\epsilon}{\Delta^2}
-\frac{\epsilon^2}{2\Delta^2}\right)\right] }\, ,
$$ 
and the dominant domains in the integration variables are given  
by 
$$
T\lsim \frac{1}{l}\, \frac{\Delta^2}{M_B}\, , \; \;  
\epsilon \lsim  \frac{1}{l}\Delta \, .
$$
All expressions then simplify, 
and the two integrations 
decouple 
$$
\gamma(n)=
\frac{1}{8\pi}
\frac{\Delta^{8+2l} }{M_B^3}\times 
$$
$$
\left\{\; \int_0^\infty {\rm d}\tau \tau^{\frac{1}{2}}
\left(\tau + 2\tau_0 \right)^{\frac{1}{2}} 
\left(\tau + \tau_0 \right)
{\rm e}^{-(l+1)\tau } 
\int_0^\infty {{\rm d}\epsilon}\, 
{\rm e}^{-2(l+1)\frac{\epsilon}{\Delta}}\, 
w_1\left(\epsilon , \: \frac{\tau\Delta^2}{2M_B}\right) 
\right. \; +
$$
\beq 
\left. 
\frac{\Delta^2}{12M_B^2}
\int_0^\infty\,{\rm d}\tau \tau^{\frac{3}{2}}
\left(\tau +2\tau_0\right) ^{\frac{3}{2}}
\left(\tau +\tau_0\right)
{\rm e}^{-l\tau }
\int_0^\infty\,{{\rm d}\epsilon}\;
{\rm e}^{-2l\frac{\epsilon}{\Delta}}
w_2\left( \epsilon, \:\frac{\tau\Delta^2}{2M_B}\right) \; \right\}
\label{4}
\eeq
where 
$$
\tau \equiv \frac{2M_BT }{\Delta^2}\; , \; \;\;
\tau_0 \equiv \frac{2 M_D M_B}{\Delta^2}\: .
$$

The integrals over the excitation energy $\epsilon$ 
\beq 
J_i(T;\, \sigma
)\; \equiv \;\frac{1}{2\pi} \,\int_0^\infty \;{{\rm d}\epsilon}\: {\rm
  e}\,^{-\frac{\epsilon}{\sigma}}\, w_i (\epsilon ,\, T)
\label{borel}
\eeq are a combination of the sum rule considered in
Refs.~\cite{vcb,optical}.  They are related to the Borel transform of
the forward scattering amplitude of the weak current off the $B$
meson. The quantity $\sigma$ is the Borel parameter, and expanding
$J_i(T;\, \sigma )$ in $1/\sigma$ gives us the first, second, and so on
sum rules of Ref. \cite{optical}.  The Borelized version has an
advantage, however, of providing an upper cut off in $\epsilon$ in a
natural way.  This is important in the analysis of the perturbative
corrections; the second argument of $J$ defines the normalization point.

One readily reads off from \eq{4} that only a fraction $\sim 1/n$ of
the total energy release is fed into the final hadronic system.  The
three-momentum carried by the final state hadrons is likewise small,
namely of order $\sqrt{m_c\Delta /n}$ for $b\ra c$ and $\Delta /n$ 
for
$b\ra u$, respectively; the latter case is obtained from the former by
setting $M_D = 0$. This implies that by evaluating the various
quantities at a scale $\sim \Delta /n$ rather than $\Delta$ one
includes the potentially large higher order corrections. These
conclusions can be made more transparent by considering two 
limiting cases.

\vspace{0.3cm}

(a) If $m_c$ stays fixed, the limit $n\ra \infty$ leads to the SV
regime, as is expressed by $v \sim (m_b-m_c)/nm_c \ll 1 $.  
Neglecting
$\tau$ compared to $\tau_0$, one gets
$$ 
\gamma(n) \;= \; \Delta^{2l+5} \left(\frac{M_D}{M_B} \right)^{3/2}
\int_0^\infty\,{\rm d}\tau \tau^{1/2} {\rm e}^{-(l+1)\tau}
J_1\left(\frac{\tau\Delta^2}{2M_B};\; \frac{\Delta}{2(l+1)} \right) \;+
$$ 
\beq 
\frac{1}{3}\Delta^{2l+5} \left(\frac{M_D}{M_B} \right)^{5/2}
\int_0^\infty\,{\rm d}\tau \tau^{3/2} {\rm e}^{-l\tau}
J_2\left(\frac{\tau\Delta^2}{2M_B}; \; \frac{\Delta}{2l} \right)\;.
\label{5}
\eeq 
The behavior of $\Gamma_{\rm sl}$ at $n\ra \infty$ then is
determined by the integrals $J_{1,2}$ near zero argument. The 
second
term is clearly subleading because the weight function contains an
extra power of $\tau$. It is not difficult to see that the axial
current contribution is dominant, and one obtains for the reduced
width \beq \gamma(n)|_{n \ra \infty} \simeq \; \Delta^n \sqrt{2\pi}
\left(\frac{M_D}{M_B}\right)^{3/2} \frac{1}{n^{3/2}}\; \xi_A\!\left(
  \frac{\Delta}{n}\right)
\label{6}
\eeq where $\xi_A(\mu)= 1+{\cal O}(\as (\mu ))$ is the coefficient
function of the unit operator in the first sum rule for the axial
current at zero recoil (for further details see Ref.~\cite{optical}).
To the degree of accuracy pursued here (power corrections are 
switched
off so far) one has $\xi_A \simeq \eta_A^2$ where the factor 
$\eta_A$
incorporates the perturbative corrections to the axial current at zero
recoil.\footnote{The quantity $\eta_A^2\equiv \lim_{\mu\rightarrow 
0}
  \xi_A^{\rm pert}(\mu)$ is ill-defined once power corrections are
  addressed; at the same time $\xi_A(\mu)$ is even then a well 
defined
  quantity provided that $\mu \gg \Lambda_{\rm QCD}$.  Yet, to any
  finite order in perturbation theory, one can work with $\eta_A$.}
Eq.~(\ref{6}) provides a decent approximation to the true width even
for $n=5$.  The vector current contribution is subleading; to take it
into account at $n=5$ we need a refined expansion to be described in
Appendix.

We note that the leading-$n$ expansion of the width yields
$(M_B-M_D)^n$ which {\em differs} from
$(\overline{m}_b-\overline{m}_c)^n$ 
by terms $(n\,\as)^k$ that we
target, but coincides to this accuracy with
$\left(m_b(\mu)-m_c(\mu)\right)^n$ if $\mu \lsim m_b/n$,
$$ m_b(\mu)-m_c(\mu) \simeq ({M}_B-{M}_D) + (m_b-m_c)
\frac{\mu_\pi^2(\mu)-\mu_G^2(\mu)}{2m_bm_c}+\,...
$$ 
\beq \simeq ({M}_B-{M}_D)\,\left(1-\frac{4\as}{3\pi}
\frac{\mu^2}{2m_bm_c}+\,...\,\right) \, .
\label{10a}
\eeq 
In the last line we used the perturbative expression for
$\mu_\pi^2(\mu)-\mu_G^2(\mu)$.
\vspace*{0.3cm}

(b) Another case of interest is $m_q \ll m_b/n$, when the final-state
quark is ultrarelativistic. In this case the axial and vector currents
contribute equally. The reduced width now takes the form 
\beq
\gamma(n)= \frac{1}{4}M_B^n \int_0^\infty\,{\rm d}\tau \, \tau^2 
\,{\rm e}\,^{-(l+1)\tau} J_1 \left(\frac{\tau M_B}{2};\; 
\frac{M_B}{2(l+1)}
\right)\, .
\label{7}
\eeq 
The quantity $J_1(\tau M_B/2)$ stays finite (unity at the tree
level) at $\tau \ra 0$. The second structure function again formally
yields only subleading in $1/n$ contributions. 
Thus 
\beq 
\gamma_n(b\ra u)|_{n\ra \infty} \;\simeq \; M_B^n \,\frac{4}{n^3} \, 
\xi_u(M_B/n)\: , 
\eeq
where 
\beq \xi_u(n)\;=\;J_1 \left( \frac{M_B}{n}; \:\frac{M_B}{n} \right)\;
, \;\;\; \xi_u^{\rm tree}=1\; .
\label{8}
\eeq 
This width decreases faster with $n$ than for massive quarks 
due
to the higher power of $|\vec q\,| \sim m_b/n$. This underlies the
fact that the numerical factor in $\Gamma_{\rm sl}$ in front of
$\Delta^5$ is much smaller for $b\ra u$ than in the SV limit, namely,
$1/192$ vs.  $1/15$.  For the same reason the simple expansion
described above provides a poor approximation for $b\ra u$ when
evaluated at $l=0$, i.e. $n=5$.  \vspace*{0.3cm}

(c) One can consider the third large-$n$ regime when $m_c/m_b 
\sim
1/n$.  Although it appears to be most close to the actual situation,
we do not dwell on it here: the analysis goes in the very same way,
but expressions are more cumbersome since one has to use the full
relativistic expression for the kinetic energy of the $D$ meson. No
specific new elements appear in this case.

\vspace*{0.3cm} (d) Let us discuss now the normalization point for
quark masses at large $n$.  Eqs.~(\ref{6}), (\ref{8}) show the
kinematically-generated dependence on the masses. The non-trivial
hadronic dynamics are encoded in the factors $\xi_A$ and $\xi_u$. 
Our
focus is the heavy quark masses.  In the SV case (a) the dependence 
on
them is trivial as long as the low-scale masses are employed.  In the
case of the $b\ra u$ transitions we showed in Sect. 2.1 that the
factor $\xi_u(M_B/n)$ actually converted $M_B^n$ into $m_b (\mu 
)^n$,
and, thus, effectively established the normalization point for the
latter, $\mu \lsim m_b/n$.  Using the effective running mass $m_b (\mu 
\sim
m_b/n)$ does not incorporate, however, the domain of the gluon 
momenta
above $\sim m_b/n$.  This contribution has to be explicitly 
included
in the perturbative corrections.  It is not enhanced by powers of $n$.
Physically, it is nothing but a statement that for $\Gamma(b\ra u)$
the proper normalization scale {\em for masses} is $\mu\sim 
m_b/n$.

We hasten to emphasize that the statement above is {\em not} a
question of normalization of $\as$ used in the perturbative
calculations. One can use $\as$ at any scale to evaluate $m_b(\mu)$ 
as
long as this computation has enough accuracy. The fact that using
inappropriate $\as$ can lead to an apparent instability in 
$m_b(\mu)$,
is foreign to the evaluation of the width.

\vspace*{0.3cm}

Let us summarize the main features which are inferred for inclusive
widths by analyzing the straightforward large-$n$ expansion 
introduced
via Eq.~(\ref{4}).  The integral over the lepton phase space carries
the main dependence on $n$. The QCD corrections -- the real
theoretical challenge -- are only indirectly sensitive to $n$: for the
kinematics determines the energy and momentum scales at which 
the
hadronic part has to be evaluated.
 
The characteristic momentum scale $\mu$ for the inclusive decay 
width
is smaller than the naive guess $\mu\sim m_b$. In the large $n$ 
limit
it scales like $m_b/n$. This momentum defines the relevant domain 
of
integration of the structure function in $q_0$, i.e. the scale at
which the forward transition amplitude appears in the total decay 
rate.
In particular, by evaluating the quark masses that determine the 
phase
space at this scale $\sim m_b/n$ one eliminates the strongest
dependence of the radiative corrections on $n$.

The expansion in $n$ derived from \eq{4} allows a transparent
discussion of the underlying physics. Unfortunately it yields, as
already stated, a decent numerical approximation only for very large
values of $n$; i.e., for $n=5$ the non-leading terms are still
significant. Not all those terms are dominated by kinematical effects,
and their treatment poses non-trivial problems. A more refined
expansion can be developed that effectively includes the
kinematics-related subleading contributions and leads to a good
approximation already for $n=5$.  This treatment, however, is rather
cumbersome and less transparent.  It will be briefly described in the
Appendix. The purpose is only to demonstrate that the essential
features of our expansion leading to the qualitative picture we rely
on -- that the characteristic momentum scale is essentially lower 
than
$m_b$ -- hold already for the actual case $n=5$. The reader who is
ready to accept this assertion, can skip the technicalities of the
refined $1/n$ expansion in the Appendix.

\section{Applications} 

We now briefly consider the consequences of the $1/n$ expansion 
for a
few problems of interest.

\subsection{The extended SV limit}

In many respects one observes that the inclusive $b\ra c$ decays 
seem
to lie relatively close to the SV limit. Most generally, the various
characteristics of the decay depend on the ratio $m_c^2/m_b^2$;
although it is rather small, the actual characteristics often do not
differ much from the case where it approaches $1$, when the 
average
velocity of the final-state $c$-quark was small.  The proximity of the
inclusive $b\ra c$ decays to the SV limit has two aspects.  First, it
is obvious that the nonperturbative effects work to suppress the
effective velocity of the final state hadrons; the most obvious
changes are kinematical replacement $m_b \ra M_B$ and $m_c \ra
M_{D,D^*,D^{**}...}$ when passing from quarks to actual hadrons. The
impact of increasing the mass is much more effective for the
final-state charm than for the initial-state beauty. This decreases
the velocities of the final state hadrons significantly at
$\Delta=m_b-m_c\simeq 3.5\GeV$.  Yet this simple effect would not 
be
numerically large enough were it not considerably enhanced by the
properties of the lepton phase space, which can be easily seen
comparing it, for example, with the one for semileptonic decays at
fixed $q^2=0$.

On the other hand, without any nonperturbative effects, theoretical
expressions at the purely parton level are known to favor the SV
kinematics even for actual quark masses.  The simplest illustration is
provided by the tree-level phase space, $z_0$: \beq z_0(m_b,m_c) 
\;=\;
m_b^5\left(1-8\frac{m_c^2}{m_b^2}-
12\frac{m_c^4}{m_b^4}\log{\frac{m_c^2}{m_b^2}}+ 
8\frac{m_c^6}{m_b^6}-
\frac{m_c^8}{m_b^8}\right) \;\;.
\label{a1}
\eeq It is most instructive to analyze the sensitivity of this
expression to $m_b$ and $\Delta=m_b - m_c$ rather than $m_b$ and
$m_c$, as expressed through \beq \kappa_\Delta\;\equiv\;
\frac{\Delta}{z_0} \frac{\partial z_0(m_b,\,\Delta)}{\partial \Delta}
\;,\;\;\; \kappa_b \;\equiv \; \frac{m_b}{z_0} \frac{\partial
  z_0(m_b,\,\Delta)}{\partial m_b}\; ,
\label{a2}
\eeq where
$$ \kappa_\Delta + \kappa_b = 5 \;.
$$ In the light quark limit -- $m_c^2/m_b^2\ra 0$ -- one has
$z_0|_{\rm light} = m_b^5\left(1-{\cal
  O}\left(\frac{m_c^2}{m_b^2}\right)\right)$ and, thus \beq
\kappa_\Delta|_{\rm light}=0 \; \; , \; \; \kappa_b|_{\rm light}=5 \;
.  \eeq In the SV limit, on the other hand, one finds \beq z_0|_{\rm
  SV} \simeq \frac{64}{5} (m_b - m_c )^5 \;\;.  \eeq Therefore, the SV
limit is characterized by \beq \kappa_\Delta|_{\rm SV}=5 \; \; , \; \;
\kappa_b|_{\rm SV}=0 \eeq For actual quark mass values -- $m_c 
/m_b
\simeq 0.28$ -- one finds \beq \kappa_\Delta\simeq 3 \; \; , \; \;
\kappa_b \simeq 2 \eeq i.e., even for $m_c^2/m_b^2 \simeq 0.08 \ll 
1$
one is still closer to the SV than the light-quark limit. The
``half-way'' point -- $\kappa_\Delta = \kappa_b = 2.5$ -- lies at
$m_c^2/m_b^2 \simeq 0.05$! This is a consequence of the large
parameter $n=5$. A similar pattern persists also for the low-order
perturbative corrections \cite{look,upset}, and on the
level of the $1/m^2$ power corrections \cite{buv,dpf,prl}.

The relevance of the SV approximation in the actual inclusive decays
thus finds a rational explanation in the $1/n$ expansion.

\subsection{Resummation of large perturbative corrections}

In computing the perturbative corrections for large $n$ one can
potentially encounter large $n$-dependent corrections of the type
$(n\as/\pi)^k$, which makes them sizable for the actual case $n=5$.
The related practical concern of certain numerical ambiguity of the
one-loop calculations of the widths was raised in Ref.~\cite{bn}.
Fortunately, the leading subseries of these terms can readily be
summed up. The summation essentially amounts to using the quark 
masses
normalized at a low scale, $\mu\sim (m_b,\Delta)/n$.  As long as one
does not go beyond a relatively low order in $\alpha_s$ one can 
ignore
renormalon divergences and simply use the ``$k$-th order pole 
mass''.
Quite obviously, the terms $\sim(n\as/\pi)^k$ appear only if one 
uses
deep Euclidean masses. If the result is expressed in terms of
low-scale Euclidean masses (i.e. those normalized at $\mu \sim
\Delta/n$ or $\mu$ = several units $\times\Lambda_{\rm QCD}$) the
terms $\sim(n\as/\pi)^k$ enter with coefficients proportional to
$\mu/\Delta$ and, therefore, contribute only on the subleading level.
                                               
The above assertion is most easily inferred by applying Eq.~(\ref{4})
to $w_i$ computed through order $k$ (with $k<n$); they are
self-manifest in Eqs.~(\ref{6}) and (\ref{8}).  Upon using the
thresholds in the perturbative $w_i$ to define $m_b$ and $m_c$, the
moments of $w_i$ are smooth and independent of $n$ in the scaling
limit $m,\,\Delta\sim n$, which ensures the absence of the leading
power of $n$. The perturbative thresholds, on the other hand, are
given just by the pole masses as they appear in the perturbation
theory to the considered order.

It is worth clarifying why these terms emerge if one uses different
masses.  The reason is that the moments of the structure functions 
(or
more general weighted averages in the `refined' expansion)
intrinsically contain the reference point for the energy $q_0$ and
$q^2_{\rm max}$. At $q^2 > q^2_{\rm max}$ the integral over $q_0$
merely vanishes whereas for $q^2 < q^2_{\rm max}$ it is unity plus
corrections. A similar qualification applies to evaluation of the
integral over the energy $q_0$ itself. These properties which single
out the proper `on-shell' masses, were tacitly assumed in carrying 
out
the expansion in $1/n$.

On the formal side, if one uses a mass other than the perturbative
pole mass, the perturbative $w_i$ contain singular terms of the type
\beq \delta(\epsilon-\delta m)= \delta(\epsilon) +
\sum_{k=1}^{\infty}\, (-\delta m)^k \frac{\delta^{(k)}(\epsilon)}{k!}
\;\;.
\label{d1}
\eeq where $\delta m \sim \as\cdot \Delta$ is the residual shift in
the energy release. (A similar shift in the argument of the
step-functions emerges for the perturbative continuum 
contributions.)
Each derivative of the $\de$-function generates, for example, a
power of $n$ in the integral over $\epsilon$ in the representation
(\ref{4}).  For $b\ra u$ widths considered in detail in Sect.~2, using
$\overline{m}_b(m_b)$ would yield $-\delta m \simeq 
\frac{4\as}{3\pi}
m_b$ leading to the series $\left(\frac{4n\as}{3\pi}\right)^k/k!$
which sums into ${\rm e}\,^{n 4\as/3\pi}$ and converts
$\overline{m}_b(m_b)^n$ into $m_b^n(\mu)$ with $\mu \ll m_b$.

A more detailed analysis can be based on the refined $1/n$ 
expansion
described in the Appendix, Eqs.~(\ref{10})--(\ref{11b}); it will be
presented elsewhere. For our purposes here it is enough to note that
using masses normalized at a scale $\sim \Delta/n$ does not 
generate
terms $\sim(n\as/\pi)^k$. \footnote{It is important that here we
  consider large $n$ for a given order in the perturbative expansion.}

A qualifying comment is in order. Including higher order terms in 
the
expansions in $1/n$ and in $\as$, one cannot find a universal scale
most appropriate for {\em all} perturbative corrections
simultaneously, including the normalization of the strong coupling.
Even upon resumming all $n$-dependent effects one would end up 
with
the ordinary corrections to the structure functions, e.g. those
describing the short-distance renormalization of the vector and axial
currents. The relevant scale for such effects is obviously $\sim m_b$;
we are not concerned about them here since they are not enhanced 
by the factor $n$ and, thus, not large.

\subsection{Power corrections and duality in inclusive widths}

Using $n$ as an expansion parameter one can estimate the 
importance of
the higher order nonperturbative corrections by summing up the 
leading
terms in $n$. The simple estimates show, however, that these effects
do not exceed a percent level, and thus are not of practical
interest.\footnote{The $b \ra c\, \tau \nu$ decays represent a special
  case: due to the sizable $\tau$ mass the energy release is smaller
  than in $b \ra c\, \mu \nu$ and $b \ra c \,{\rm e} \nu$ and a SV
  scenario is realized in a rather manifest way. The $n$-enhanced
  nonperturbative corrections can then reach the ten percent level,
  but are readily resummed using the exact meson masses in the 
phase
  space factors.} It is still instructive to mention a qualitative
feature drawn from this analysis.

The characteristic momentum scale decreases for large $n$; at
$\Delta/n \sim \Lam$ one is not in the short-distance regime 
anymore.
Correspondingly, the overall theoretical accuracy of the calculations
naturally seems to deteriorate with increasing $n$ (and/or increasing
$m_c$ up to $m_b$). However, using the sum rules derived in
\cite{optical} it is not difficult to show that even in the limit
$n\ra \infty$ the width is defined by short-distance dynamics
unambiguously up to $1/m_Q^2$ terms. Moreover, in this limit a
straightforward resummation of the leading nonperturbative effects
yields the width in terms of the zero-recoil formfactor $|F_{D^*}|^2$
($|F_{D}|^2$) and the physical masses $M_B$ and $M_{D^{(*)}}$, 
without
additional uncertainties. Thus, in the {\em worst} limit for inclusive
calculations, the theoretical computation of the width merely reduces
to the calculation of the exclusive $B\ra D^*$ transition, the best
place for application of the heavy quark symmetry \cite{SV,IW}.

The large-$n$ arguments may seem to suggest that the
width can depend on the meson masses rather than on the quark 
ones, in
contradiction to the OPE.  In fact, since in the $b\ra c$ transitions
one encounters the SV limit when $n\ra \infty$, the leading term is
given by $(M_B-M_D)^n$ which differs from $(m_b-m_c)^n$ only by 
${\cal O} \left(\Lam^2/m_Q^2\right)$, in accordance with the OPE. The
problem can emerge in the next-to-leading order in $1/n$. These
corrections can be readily written up in terms of the transition
formfactors at small velocity transfer, and take the form of the sums
of the transition probabilities which one encounters in the small
velocity sum rules \cite{optical}. The latter ensure that the terms
$\sim \Lam/m_Q$ are absent. By the same token, using the third 
sum
rule, one observes that the leading-$n$ term, $n \mu_\pi^2/m_Q^2$, 
is
absent from the width although it is obviously present in
$(M_B-M_D)^n$ (we assume that the quark masses are fixed as the
input parameters).  Similar properties hold to all orders in $1/n$,
however they are not very obvious in this expansion.

If the infrared part of the quark masses cancels in the widths, why is
it not the short-distance masses like $\overline{\rm MS}$ that 
emerge?
The answer is clear: the excited final states in the width enter with
the weight $\sim {\rm exp}{(-n\epsilon/\Delta)}$, and the Coulomb 
part
of the mass cancels in the width only up to momenta $\sim 
\Delta/n$.

The above consideration only interprets the known OPE results.  The
general question at which scale duality is violated is more
instructive.  We immediately see that, generically, the characteristic
momenta are given by the scale $m_Q/n$ rather than by $m_Q$.  For
example, considering the calculation of the relevant forward
transition amplitude for $b\ra u$ in coordinate space \cite{buv,dual},
we have \beq T\; \sim \; 1/x^{n+4}\, .
\label{c1}
\eeq Its Fourier transform $\sim p^n \log{p^2}$ ( $+$ terms analytic
in $p^2$) is saturated at $|x| \sim r_0 \sim p/n$, corresponding to a
time separation $t_0 \sim n/m_Q$. In the case of the massive final
state quarks one has $t_0 \sim n /\Delta$.

In the semileptonic $b\ra c$ decays one may naively conclude that 
the
scale is too low. However, here the heavy quark symmetry ensures
vanishing of all leading corrections even at $\Delta \ra 0$, i.e. in
the SV limit; this applies as well to the ESV limit. The nontrivial
corrections appear only at the $1/\Delta^2$ or $n/(m\Delta)$ level 
and
they are still small.

The situation is, in principle, different in the nonleptonic decays,
where the color flow can be twisted (instead of the color flow from
$b$ to $c$, it can flow from $b$ to $d$). In the nonleptonic $b\ra
c\,\bar u d$ decay, with $\Delta \simeq 3.5\GeV$, one is only
marginally in the asymptotic freedom domain and {\em a priori}
significant nonperturbative corrections as well as noticeable
violations of duality could be expected. In this respect it is more
surprising that so far no prominent effects have been identified on
the theoretical side \cite{dual}.\footnote{A priori the effect of the
chromomagnetic interaction in order $1/m_Q^2$ could have been
significant, but it gets suppressed due to the specific chiral and
color structure of weak decay vertices \cite{buv}, properties that
are external to QCD. } We note that a particular model for the
violation of duality in the inclusive decays suggested in Ref.
\cite{dual} explicitly obeys the large-$n$ scaling above. The
importance of the duality-violating terms is governed by the ratio
$\Delta/n$, 
\beq 
\frac{\Gamma_{\rm dual.viol.}}{\Gamma_{\rm  tree}}
\;\sim \; {\rm e}\, ^{-\Delta \cdot \rho \left(1-\frac{n}{\Delta\cdot \rho}
\ln {\frac{n}{\Delta\cdot \rho} } \right)}\;\;,
\label{c2}
\eeq 
with $\rho$ being a hadronic scale parameter. The corrections
blow up when $\Delta/n \sim \mu_{\rm had}$. We hasten to remind,
however, that this consideration is only qualitative, since we do not
distinguish between, say $n=5$ and $n-2=3$. The arguments based 
on the
refined expansion in the ESV case suggest that the momentum scale 
is
governed by an effectively larger energy than literally given by
$\Delta/n$ with $n=5$.

\section{Low-scale heavy quark mass; renormalized effective 
operators}

Addressing nonperturbative power corrections in $b$ decays via
Wilson's OPE one must be able to define an effective nonrelativistic
theory of a heavy quark normalized at a scale $\mu \ll m_Q$; the 
heavy
quark masses in this theory (which enter, for example, equations of
motion of the $Q$ fields) are low-scale (Euclidean) masses 
$m_Q(\mu)$.
Moreover, even purely perturbative calculations benefit from using
them, as was shown above: it allows one to solve two problems
simultaneously -- resum the leading $n$-enhanced terms and avoid 
large
but irrelevant IR-renormalon--related corrections.  For the
perturbative calculations {\em per se}, therefore, the question arises
of what is the proper and convenient definition of a low-scale
Euclidean mass?  Of course, one can always define it through an
off-shell quark propagator. Generically this will lead to a mass
parameter which is not gauge invariant; this feature, though not 
being
a stumbling block, is often viewed as an inconvenience.

Usually, one works with the mass defined in the $\overline{\rm 
MS}$
scheme, which is computationally convenient. However, due to the
well-known unphysical features of this scheme it is totally
meaningless at the low normalization point $\mu$ = several units
$\times \Lambda_{\rm QCD}$.  For example, if we formally write the
$\overline{\rm MS}$ mass at this point \beq m_Q^{\overline{\rm
    MS}}(\mu) \;\simeq \;m_Q^{\overline{\rm MS}}(m_Q) \,
\left(1+\frac{2\as}{\pi}\ln\frac{m_Q}{\mu}\right)
\label{5.0}
\eeq we get large logarithms $\ln m_b/\mu$ which do not reflect 
any
physics and are not there under any reasonable definition of the 
heavy
quark mass.

Thus, one must find another definition. We suggest the most physical
one motivated by the spirit of the Wilson's OPE, i.e. the mass which
would be seen as a `bare' quark mass in the completely defined
effective low-energy theory.

The basic idea is to follow the way suggested in \re{optical}, and is
based on the consideration of the heavy flavor transitions in the SV
kinematics, when only the leading term in the recoil velocity, $\sim
\vec{v}^{\,2}$, is kept. Note that in Subsection 2.1 we have used a 
similar
purpose a different kinematical limit when the final particle is 
ultrarelativistic. 

Let us consider the heavy-quark transition $Q\ra Q'$ induced by 
some
current $J_{QQ'}=\bar Q' \Gamma Q$ in the limit when $m_{Q,Q'} \ra
\infty$ and $\vec{v}=\vec q /m_{Q'}\ll 1$ is fixed. The Lorentz
structure $\Gamma$ of the current can be arbitrary with the only
condition that $J_{QQ'}$ has a nonvanishing tree-level nonrelativistic
limit at $\vec{v}=0$. The simplest choice is also to consider $Q'=Q$
(which, for brevity of notation, is assumed in what follows). Then one
has \cite{optical} 
\beq 
\La(\mu)\; \equiv \; \lim_{m_Q \ra
  \infty}\;\left[ M_{H_Q}-m_Q(\mu)\right]\; = 
\; \lim_{\vec v \ra 0}\,
\lim_{m_Q \ra \infty} \frac{2}{\vec v^{\,2}} \, \frac{\int_0^\mu \;
  {{\rm d}\epsilon}\,\epsilon\, w(\epsilon,\,m_{Q} v)} {\int_0^\mu
  \;{{\rm d}\epsilon}\, w(\epsilon,\,m_{Q} v)} \;\;,
\label{5.1}
\eeq 
where $w(\epsilon,\,|\vec{q}\,|)$ is the structure function of the
hadron (the probability to excite the state with the mass
$M_{H_Q}+\epsilon$ for a given spatial momentum transfer 
$\vec{q}\,$); 
$M_{H_Q}$ is the mass of the initial hadron.

The OPE and the factorization ensures that

a) For the given initial hadron the above $\La(\mu)$ does not 
depend
on the chosen weak current.

b) The value $m_Q(\mu)=M_{H_Q}-\La(\mu)$ does not depend on 
the choice
of the hadron $H_Q$. This holds if $\mu$ is taken above the onset of
duality.  

To define perturbatively the running heavy quark mass one merely
considers \eq{5.1} in the perturbation theory, to a given order:
\beq
m_Q(\mu)\; =\;\left[ M_{H_Q}\right]_{\rm pert}-
\left[\La(\mu)\right]_{\rm pert}\;\;.
\label{5.1.1}
\eeq
In the perturbation theory  $H_Q$ is the
quasifree heavy quark state and $M_{H_Q}$ is the pole mass 
$m_Q^{\rm pole}$ to
this order; $w(\epsilon)$ at $\epsilon >0$ is a perturbative 
probability to 
produce $Q$ and a number of gluons in the final state which belongs 
to the
perturbative continuum. It is a continuous function (plus 
$\delta(\epsilon)$ at
$\epsilon =0$). Both numerator and denominator in the definition of 
$\La$ are
finite. All quantities are observables and, therefore, are manifestly 
gauge-invariant. For example, to the first order 
$$
\left[\La(\mu)\right]_{\rm pert}\;=\;
\frac{16}{9}\frac{\as}{\pi}\mu\;,\;\;\;
m_Q(\mu)\; =\; m_Q^{\rm pole_{1-loop}} - \frac{16}{9}\frac{\as}{\pi}
\mu\;\;,
$$
\beq
m_Q(\mu)\;=\; m_Q^{\overline{\rm 
MS}}(m_Q)\left(1+\frac{4}{3}\frac{\as}{\pi}- 
\frac{16}{9}\frac{\as}{\pi}\frac{\mu}{m_Q}\right)\;\;.
\label{5.3}
\eeq
The IR renormalon-related problems encountered in calculation of
$m_Q^{\rm pole}$ cancel in Eq.~(\ref{5.1.1}) against the same
contribution entering $\La(\mu)$ in Eq.~(\ref{5.1}).

In a similar way one can define matrix elements of renormalized at 
the 
point $\mu$ heavy quark operators \cite{optical,pert,rev}. The 
zeroth
moment of the structure function is proportional to 
$\matel{H_Q}{\bar Q
Q}{H_Q}/(2M_{H_Q})$ and to the short-distance renormalization 
factor. In
the limit $m_Q\ra \infty$ this matrix element equals unity up to 
terms
$\mu^2/m_Q^2$. Then it is convenient to define matrix elements  of 
higher dimension operators in terms of the ratios (for which all 
normalization factors go away):
\beq
\mu_\pi^2(\mu)\,=\,\frac{\matel{H_Q}{\bar Q (i\vec{D}\,)^2 Q}{H_Q}_\mu}
{\matel{H_Q}{\bar Q Q}{H_Q}_\mu} 
\;=\;
\lim_{\vec v \ra 0}\,
\lim_{m_Q \ra \infty}
\frac{3}{{\vec v}^{\,2}}
\frac{ 
\int_0^\mu \;{{\rm d}\epsilon}\,\epsilon^2\, w(\epsilon,\,m_{Q} v)}
{\int_0^\mu \; {{\rm d}\epsilon}\, w(\epsilon,\,m_{Q} v)}
\;\;,
\label{5.4}
\eeq
\beq
\rho_D^3(\mu)\,=\, -\frac{1}{2}
\frac{\matel{H_Q}{\bar Q\,( \vec D \vec E)\, Q}{H_Q}_\mu}
{\matel{H_Q}{\bar Q Q}{H_Q}_\mu }
 \;=\;
\lim_{\vec v \ra 0}\,
\lim_{m_Q \ra \infty}
\frac{3}{{\vec v}^{\,2}}
\frac{ 
\int_0^\mu \;{{\rm d}\epsilon}\,\epsilon^3 \, w(\epsilon,\,m_{Q} v)}
{\int_0^\mu \; {{\rm d}\epsilon}\, w(\epsilon,\,m_{Q} v)}
\;\;,
\label{5.5}
\eeq
and so on. All these objects are well-defined, gauge-invariant and do 
not
depend on the weak current probe.

Returning to the heavy quark mass, the literal definition given above
based on \eq{5.1.1} is most suitable for $\mu \ll m_Q$ when the 
terms
$\sim (\as/\pi) (\mu^2/{m_Q})$ are inessential. One 
can
improve it including higher-order terms in $\mu/m_Q$, if necessary, 
in
the following general way.

Consider the $1/m_Q$ expansion of the heavy hadron mass
\beq
\left(M_{H_Q}\right)_{\rm spin-av}\; =\; m_Q(\mu) + \La(\mu)+ 
\frac{\mu_{\pi}^2(\mu)}{2 m_Q(\mu)} +
\frac{\rho_D^3(\mu)-\rho^3(\mu)}{4m_Q^2(\mu)}+\;...\;\;;
\label{5.5a}
\eeq
We took the average hadronic mass over the heavy quark spin 
multiplets
here. 
Besides $\mu_{\pi}^2$ and $\rho_D^3$ defined above the new 
quantity 
 $\rho^3$ enters the $1/m_Q^2$ term. It is a non-local correlator of 
two operators $\bar Q (\vec\sigma
i\vec D)^2 Q$ defined in Eq.~(24) of \cite{optical}.
 We apply the relation (\ref{5.5a}) to the perturbative states and get 
\beq
m_Q(\mu)\; =\; m_{Q}^{\rm pole}-\left[\La(\mu)\right]_{\rm pert}-
\frac{\left[\mu_{\pi}^2(\mu)\right]_{\rm pert}}{2 m_Q(\mu)} -
\frac{\left[\rho_D^3(\mu)\right]_{\rm pert}-
\left[\rho^3(\mu)\right]_{\rm pert}}{4m_Q^2(\mu)}-\;...\;\;.
\label{5.6}
\eeq
For example, through order $1/m_Q$ one has 
\beq
\left[\mu_{\pi}^2(\mu)\right]_{\rm pert}\simeq \frac{4\as}{3\pi} 
\mu^2\;.
\label{5.7}
\eeq 
To renormalize the non-local
correlators entering the definition of $\rho^3$,  one
represents them as a QM sum over the intermediate states and cuts 
it off at $E_n \ge M_{H_Q} +\mu$. 
For example, $\left[\rho^3(\mu)\right]_{\rm pert} \simeq  (8\as/9\pi) 
\mu^3.$
One subtlety is worth mentioning here:
including higher orders in $1/m_Q$ and going beyond one-loop
perturbative calculations one must be careful calculating the 
coefficient
functions in the expansion Eq.~(\ref{5.5a}).

The general definition above is very transparent physically and most
close to the Wilson's procedure for a theory in {\em Minkowski} 
space.
It corresponds to considering the QM nonrelativistic system of slow
heavy flavor hadrons and literally integrating out the modes with
$\omega = E-m_Q > \mu$. Considering the SV limit and keeping only
terms linear in $\vec{v}^{\,2}$ is important: in this case no problem
arises with choosing the reference energy in $\epsilon$ or in 
imposing
the cutoff (energy vs. invariant mass, etc.) -- they are all 
equivalent in this approximation; the differences appear starting
terms $\sim \vec{v}^{\,4}$. The conceptual details will be further 
illuminated in subsequent publications.

\section{Determination of $|V_{cb}|\;$ ($|V_{ub}|$) from the
inclusive widths}

We now 
address the central phenomenological problem -- how accurately one
can extract $|V_{cb}|$ and $|V_{ub}|$ from inclusive semileptonic 
widths. ${\cal O}(\as )$ terms are known, and higher order ones are
expected to be small for $b$ decays. 
However radiative corrections to  the mass 
get magnified due to the high power 
with which the latter enters the width. This concern has been raised 
in
\cite{bn}. The effect of large $n$ indeed shows up as the series
of terms $\sim (n\as/\pi)^k$. Therefore 
one needs to sum up such terms. 
The resummation of the running $\alpha_s$ effects in the
form $\frac{\as}{\pi}\left(\frac{\beta_0}{2}\frac{\as}{\pi}\right)^k$ 
has been
carried out in \cite{bbbsl} (the term with $k=1$ in this series was earlier 
calculated in \cite{wise}); with $\beta_0/2=4.5\,$, accounting for 
the large-$n$
series seems to be at least as important. We emphasize that the 
BLM-type
\cite{BLM} improvements leave out these large-$n$ terms.

In this paper we discussed a prescription  automatically resumming  
the  
$(n\as/\pi)^k$ terms. 
This eliminates the source of the perturbative uncertainty often 
quoted
in the literature as jeopardizing the calculations of the
widths.\footnote{These uncertainties do not apply to our previous
analyses \cite{vcb,upset} where we have employed quark masses 
evaluated at a low scale 
keeping in mind the natural momentum scale inherent for the actual
inclusive widths, following the arguments under discussion. They are
intrinsic, however,  to some alternative estimates which can be found 
in the 
literature.}
Let us demonstrate this assertion in a simplified setting.

Let us consider two limiting cases, $m_q\ll m_b$ and $m_q\simeq 
m_b$. 
Dropping from the widths inessential overall factors
$G_F^2|V_{qb}|^2/(192\pi^3)$ ($\;G_F^2|V_{qb}|^2/(15\pi^3)\;$) we 
get
\beq
\Gamma \simeq \Delta^n\left(1+a_1\frac{\as}{\pi} +
a_2\left(\frac{\as}{\pi}\right)^2\;+\;...\right)\;\;,
\label{e1}
\eeq
with $a_1=-\frac{2}{3}\left(\pi^2-\frac{25}{4}\right)$ (or $a_1=-1$,
respectively) at $n=5$. As was shown before, these coefficients do 
not contain
factors that scale like $n$, 
nor $a_2$ contains $n^2$, etc., and we can neglect
them altogether. However, when one uses, say, the $\overline{\rm 
MS}$ masses,
one has 
\beq
\overline{m}_Q \;\simeq\; m_Q\left(1-\frac{4}{3}\frac{\as}{\pi}\, ,
\right)
\label{e2}
\eeq
and to the same order in $\as$ one arrives, instead, at a different 
{\em
numerical} estimate
$$
\Gamma\; = \; \Delta^n\left(1-\frac{4}{3}\frac{\as}{\pi}\right)^n 
\left(1+n\frac{4}{3}\frac{\as}{\pi}\right)\, \ra
$$
\beq 
\overline{\Delta}^{\,n} 
\left(1 -\frac{(n+1)n}{2} \left(\frac{4}{3}\frac{\as}{\pi}\right)^2
\;+\; \frac{(n+1)n(n-1)}{3} \left(\frac{4}{3}\frac{\as}{\pi}\right)^3
\;-\;...\right)\;\; ; 
\label{e3}
\eeq
$\overline{\Delta}$  is written in terms of the $\overline{\rm MS}$ 
scheme 
masses.
The expressions in \eq{e1} and \eq{e2} are equivalent 
to order $\as$, yet start to differ at order 
$\as^2$ due to $n$-enhanced terms. The size of the difference has 
been 
taken as a numerical
estimate of the impact of the higher order (non-BLM) corrections 
which then 
seemed to be
large indeed: the coefficient in front of $(\as/\pi)^2$ is more than
$25$; only such huge enhancement could lead to an uncertainty in 
the
width $\sim 15\%$  from the higher order corrections.

Since we are able to resum these terms, the width \eq{e1} has small
second (and higher order) corrections. Alternatively, using the
$\overline{\rm MS}$ masses one is bound to have large higher-order
perturbative coefficients, and their resummation returns one to using
the low-scale masses.

We, thus, conclude that the uncertainty associated with 
the higher-order
perturbative corrections discussed in Ref.~\cite{bn} is actually absent
and is an artifact of the approach relying on the $\overline{\rm MS}$
masses without a resummation of the large-$n$ effects. A similar
numerical uncertainty, in a somewhat softer form, appeared as a
difference between the ``OS'' and the ``$\overline{\rm MS}$'' 
all-order
BLM calculations of Ref.~\cite{bbbsl}; it is likewise absent if one
uses the proper low-scale masses. 

In our asymptotic treatment of the large-$n$ limit we cannot specify 
the
{\em exact} value of the normalization scale $\mu$ to be used for 
masses: 
it can equally be $0.7\GeV$ or $1.5\GeV$; the exact scale would be
a meaningless notion. We also saw that the position of the saddle 
point
varies depending on the structure function considered. All these 
nuances are 
rather unimportant in practice: the impact of changing
$\mu$ has been studied in Ref.~\cite{upset}, and it was found that 
the
values of $|V_{qb}|$ extracted from the widths vary by less than 
$\pm 1\%$  for any reasonable value of $\mu$.

The large factor of $5$ 
also enhances the sensitivity of $|V_{cb}|$ to the expectation
value of the kinetic energy operator $\mu_\pi^2$ in the currently 
used 
approach
where $m_b-m_c$ is related to $M_{B}$, $M_{B^{*}}$, $M_{D}$, 
$M_{D^{*}}$ and
$\mu_\pi^2$. This is certainly a disadvantage; however, the emerging 
dependence of $|V_{cb}|$ on $\mu_\pi^2$ is 
practically identical (but differs in sign) to the one in the 
determination 
of $|V_{cb}|$ from
the zero recoil rate $B\ra D^* \ell  \nu$  
(for details see the recent review \cite{rev}). 
We do not see, therefore, a way to eliminate the uncertainty due to
$\mu_\pi^2$ other than to extract its numerical value from the 
data. Having at hand its proper field-theoretic definition, one does 
not 
expect
significant theoretical uncertainties in it.
In
the exclusive case, unfortunately, even this would not remove the 
sizable
element of model-dependence.

The $1/n$ expansion can equally well be applied when the 
final-state quark mass is practically zero as in $b \ra u$. The 
large corrections of order $\as ^2$ that appear when $m_b$ is taken 
at the high scale $m_b$ turn out to be quite small when $m_b$ 
is evaluated at $\Delta /n \sim 1$ GeV with $\Delta$ denoting the 
energy release. Using the analysis of
Ref.~\cite{upset}\footnote{A more complete BLM calculation has been
carried out in \cite{bbbsl}; the numerical results were quoted, 
however, only for
the OS and $\overline{\rm MS}$ schemes; the latter is expected to 
suffer
from large
$n$-related corrections, whereas the OS scheme is affected by the IR
renormalons when higher-order BLM corrections are included; the 
BLM 
resummation does not cure it completely, again due to the
$n^2$-enhanced terms in the spurious $1/m^2$ corrections.} 
one can calculate the inclusive total semileptonic width 
$\Gamma (B \ra l \nu X_u)$ quite reliably in terms of 
$|V(ub)|$ and $m_b (1 {\rm GeV})$ as inferred from 
$\Upsilon$ spectroscopy. Numerically one has \cite{rev}
\beq
|V_{ub}|\; \simeq\;
0.00415\left(\frac{{\rm BR}(B\rightarrow X_u\ell\nu)}{0.0016}
\right)^{\frac{1}{2}}\left(\frac{1.55\,\rm
ps}{\tau_B}\right)^{\frac{1}{2}}\;.
\label{d40}
\eeq
The theoretical uncertainty here is smaller than the experimental 
error 
bars which can be expected in the near future.

\section{Conclusions} 

There are three dimensional parameters relevant for semileptonic 
decays of beauty hadrons: $m_b$, the energy release 
$\Delta _{bq}=m_b - m_q$ and $\Lam$. Since 
$m_b$, $\Delta _{bq} \gg \Lam$, the heavy quark expansion in 
terms of powers of $\Lam/\Delta _{bq}$  yields 
a meaningful treatment, with higher-order terms quickly fading in 
importance. It is natural then that uncertainties in the 
perturbative treatment  limit the numerical reliability 
of the theoretical predictions. It was even suggested that the
latter uncertainties are essentially beyond control.

Our analysis shows that such fears are  exaggerated. First, we 
have reminded the reader  
that the dependence of the total width on the fifth power of $m_b$ 
largely 
reflects the kinematics of the lepton pair phase space. We then used 
an expansion in $n$ to resum higher order contributions of 
kinematical 
origin by identifying unequivocally the relevant 
dynamical scales at which the quark masses have to be evaluated. 
We found 
that 
the relevant scale is not set by the energy release, but is  
lower, parametrically of order $\Delta _{bq}/n$. 

We demonstrated that the expansion in $1/n$ works reasonably well 
in
simple examples. This is not the main virtue, in our opinion. More
important is the fact that this expansion allows one to have a 
qualitative 
picture of different types of corrections based on
the scaling behavior in $n$. In
particular, it shows the emergence of new, essentially lower
scales relevant in the semileptonic decays.

Even without a dedicated analysis it is obvious that the typical scale
of the energy release in the semileptonic $b$ decays lies below 
$m_b$.
Without a free parameter at hand it is not too convincing to defend 
 say, the scale $m_b/2$ as opposed to $2m_b$.  Needless to say that 
this scale 
variation
has a noticeable impact on the final numbers. Using $n$ as an
expansion parameter the ambiguity is resolved -- the appropriate
normalization scale
for masses is even below $m_b/2$ -- and already this, quite weak, 
statement is extremely helpful  \cite{upset} in numerical estimates.

Based on the large-$n$ expansion, we arrived at a few concrete
conclusions.
\vspace*{0.25cm}

\hspace*{.1em} $\bullet$ One can resum the dominant $n^k$ terms in 
the
perturbative expansion of the inclusive widths, by merely using the 
Euclidean 
low-scale quark
masses (e.g.  normalized at the scale $\sim \Delta/n$). Therefore, the
previous calculations of the width \cite{vcb,upset} are free of the
large uncertainties noted in \cite{bn} and which were later 
claimed to be inherent to the
inclusive widths.
\vspace*{0.15cm}

\hspace*{.1em} $\bullet$ We introduced the so called  ``extended'' SV 
limit. We show that $m_c$ need not be close to $m_b$ for the SV 
regime to 
emerge in the $b\ra c$ inclusive decays. Large $n$ helps ensure this 
regime,
which gives a rationale for the relevance of the SV consideration in 
the
actual inclusive $b\ra c$ decays, both at the quark-gluon and at the 
hadronic 
levels.
\vspace*{0.2cm}

Recently, the complete second-order corrections to the zero-recoil 
formfactors have been computed \cite{czar}; the non-trivial non-BLM
parts proved to be small. The arguments based on the large-$n$ 
expansion
suggest then that the non-BLM perturbative corrections not 
computed so
far for $\Gamma_{\rm sl}(b\ra c)$ are not  large either. 

The approach suggested here is applicable to other 
problems as well. For example, we anticipate
that the second-order QED
corrections to the muon lifetime which have not been calculated so 
far,
will not show large coefficients if expressed in terms on the
muon mass normalized at the scale $\sim m_\mu/3- m_\mu/2$; 
moreover, 
using
$\alpha_{\rm em}$ at a similar scale (in the $V$ scheme) is also
advantageous -- though, clearly, it does not matter in practice in QED.
\vspace*{0.15cm}

{\em Note added:} When this paper was prepared for publication the 
full
two-loop calculation of the perturbative correction in $b\ra c 
\,\ell\nu$ at
another extreme kinematic point $q^2=0$ was completed 
\cite{czarmel}. The 
result
suggests that the non-BLM perturbative corrections to the width are 
indeed
small if one relies on the low-scale quark masses, in accord with 
our expectations.
\vspace*{0.2cm}\\ 

\noindent
{\bf Acknowledgments:} \hspace{.4em} N.U. is grateful to V.~Petrov 
for
invaluable discussions and participation in this analysis at its early
stages. We are  grateful to CERN Theory Division,  
Theory Division of MPI Werner-Heisenberg-Instit\"{u}t and to
Isaac Newton
Institute for Mathematical Sciences, Cambridge, England
where various parts of this project have been worked out, 
 for hospitality. M.S. and A.V. benefited from participation in 
 the program {\it Non-Perturbative Aspects
of Quantum Field Theory} organized by the Isaac Newton
Institute for Mathematical Sciences.

This work was supported in part by NSF under
the grant number PHY 92-13313 and by DOE under the grant 
number
DE-FG02-94ER40823.
\vspace*{1.0cm}\\

{\Large{\bf Appendix: The refined $1/n$ expansion}} 
\vspace*{0.5cm}

\noindent 
Let us start from \eq{2} expressed in terms of the 
quark masses $m_b$ and $m_c$, with 
$\Delta = m_b - m_c$, and 
introduce variables $\eta$ 
and $\omega$ defined by 
\beq
\eta=1-\frac{\sqrt{q^2}}{\Delta}\, , \; \; 
\omega=\frac{1}{\Delta}\left[ \sqrt{m_b^2 -2m_b q_0 +q^2} -m_c 
\right]
\, .
\label{9}
\eeq
Thus, $\omega$ measures the effective mass $M_X$ in the final 
hadron 
state, $\omega=(M_X - m_c)/\Delta$, (in particular, $\omega=0$ 
represents 
the free quark decay) whereas 
$\eta$ determines the difference between $q^2$ and $\Delta^2$.
Then we have
$$
\gamma(n)=
2\Delta^n \,\sqrt{\frac{\Delta}{m_b}} \left\{
\int_0^1\,{\rm d}\, (1-\eta )^{n-2}\times\right.
$$
$$
\left[\eta \left(\eta +\frac{2m_c}{\Delta}\right)
\left(1-\left(1-\frac{m_c}{2m_b}\right) 
\eta +\frac{\Delta}{4m_b}\eta ^2\right)\right]^{\frac{1}{2}} 
v_1(\eta) +
$$
\beq
\left.
 \frac{\Delta^2}{3m_b^2} \, 
\int_0^1 \,{\rm d}\eta\,\eta^{3/2}(1-\eta)^{n-4}
\left(\eta+\frac{2m_c}{\Delta}\right)^{3/2}
\left(1-\left(1-\frac{m_c}{2m_b}\right)\eta +
\frac{\Delta}{4m_b}\eta^2\right)^{3/2} v_2(\eta)
\right\}
\label{10}
\eeq
where  functions $v_i(\eta)$ are
defined through  
$$
v_1(\eta)\cdot (2m_c\eta+\eta^2)^{1/2}\cdot
\left[1-\left(1-\frac{m_c}{2m_b}\right)
\eta+\frac{\eta^2\Delta}{4m_b}\right]^{1/2}
\equiv
$$
$$
\frac{\Delta}{m_b}\int_{0}^{\eta}\,{\rm d}\omega(m_c + 
\omega\Delta) 
(\eta-\omega)^{1/2}(2m_c+\eta\Delta+\omega\Delta)^{1/2}
\times
$$
\beq
\left[1-\left(1-\frac{m_c}{2m_b}\right)
\eta-\frac{m_c}{2m_b}
\omega+\frac{\Delta(\eta^2-\omega^2)}{4m_b}\right]^{1/2}
\frac{w_1}{2\pi} \, ,
\label{11a}
\eeq

\vspace{0.2cm}
 
$$
v_2(\eta)\cdot (2m_c\eta+\eta^2\Delta)^{3/2}\cdot
\left[1-\left(1-\frac{m_c}{2m_b}\right)
\eta+\frac{\eta^2\Delta}{4m_b}\right]^{3/2}
\equiv
$$
$$
\frac{\Delta}{m_b} \int_{0}^{\eta}{\rm d}\omega\,(m_c + 
\omega\Delta)\times
$$
\beq
\left[ (\eta-\omega)
(2m_c+\eta\Delta+\omega\Delta) 
\left(1-\left(1-\frac{m_c}{2m_b}\right)
\eta-\frac{m_c}{2m_b}
\omega+\frac{\Delta(\eta^2-\omega^2)}{4m_b}
\right) \right] ^{3/2}
\frac{w_2}{2\pi}\;\;.
\label{11b}
\eeq

\vspace{0.2cm}

Equation (\ref{10}) represents an identity with two gratifying 
features. First, 
the width is expressed through the quantities $v_i$. Being 
weighted integrals of the    
structure functions $w_i$ they are smoother analytically than 
$w_i$ themselves. 
In a certain respect, $v_i$ are generalizations of the moments $I_i$, 
that 
are relevant for calculating inclusive width.

Second,
the form of \eq{10} is particularly well suited for 
deriving the large-$n$ expansion. 
One typically encounters integrals of the form 
$\int _0^1 {\rm d}\eta \eta^a(1-\eta )^k$. At $k\ra \infty$ the 
integral is saturated at $\eta_0 \simeq 1-a/(k+a)\simeq 1-a/k$, i.e.
$1-\eta _0 \ll 1$. (We loosely refer to this saturation as a
saddle-point evaluation, although it is not really a saddle point 
calculation
in its standard definition.)
This was actually the approximation used in the 
simple large-$n$ expansion. However, for 
$k=0,2$ and $a \sim 1/2 \div 3/2$ the ansatz 
$\eta _0\ll 1$ is quite poor numerically. On the other hand 
$\int _0^1 {\rm d}\eta \eta ^a(1-\eta )^k$ still has a reliable 
`saddle point' even for large $a$ and $k=0$, since the 
width of the distribution is governed by $ak/(a+k)^3$ and does 
not become large. The point only shifts somewhat upward as 
compared to the
`naive' approach. This is the idea lying behind the improved 
expansion. Its
purpose is merely to determine the essential kinematics in the 
process at hand,
and to use the hadronic averages expanded around it. Of course, the 
phase space
integrals always can be taken literally for any $n$, if necessary.

As it was with the simple 
$n$ expansion, one finds that the explicit  
dependence on $n$ is contained in the kinematical factors 
that can be treated separately from the QCD dynamics 
contained in the hadronic averages $v_{1,2}(\eta )$. 
There is some residual dependence on $n$ entering through 
the value of the scale $\eta _0$ at which 
$v_{1,2}(\eta )$ are to be evaluated (i.e., the exact shape of the 
weight
functions). 
Again, one has to treat the vector and 
axial current contributions separately for $b\ra c$ and $b\ra u$. 

In Table \ref{TABLE1} we compare the exact results with 
those obtained from the refined $1/n$ expansion for $n=5$ 
as a function of $m_c/m_b$ in the simplest setting of the tree-level 
decay
where the cumbersome factors in \eq{10} are replaced by their 
values at the
`saddle' point. 
We have used the following 
notation there. The quantities $\gamma _{1,2}^{(V,A)}$ 
denote the width factors for the vector and axial contributions 
obtained by integrating $w_1$ and $w_2$, respectively. We then 
have  
$\gamma_{\rm sl} = \gamma _1^{(A)} + \gamma _1^{(V)} + 
\gamma_2^{(A)} + \gamma _2^{(V)}$. One defines branching 
ratios normalized to the exact ($n=5$) tree level expression:
\beq
{\rm BR}_{1,2}^{(A,V)}(m_c/m_b)= 
\frac{\gamma_{1,2}^{(A,V)}(m_c/m_b)}
{\gamma_{\rm sl,\,exact}(m_c/m_b)} 
\label{12}
\eeq
and calculates them using, on the one hand, the expansion in $n$ 
evaluated at $n=5$, and on the other hand the exact $n=5$ results. 
Since the axial and vector part of $w_2$ coincide in 
the tree approximation, 
only one of them is shown in 
Table \ref{TABLE1}. 
\begin{table} 
\begin{tabular}{|c||c|c|c|c|c|c|c|} 
\hline 
$m_c/m_b $ & 0 & 0.2 & 0.3 & 0.4 & 0.6 & 0.8 & 1 \\ 
\hline 
${\rm BR}_{1,\,n=5}^{(A)}
$ & 0.223 & 0.524 & 0.462 &0.462 & 0.476 & 0.488 & 0.494 \\ 
${\rm BR}_{1,\,\rm exact}^{(A)}
$ &0.25 & 0.396 &0.434 & 0.459 & 0.486 & 0.497 & 0.5 \\
${\rm BR}_{1,\,n=5}^{(V)}
$ & 0.229 & 0.093 & 0.060 & 0.0374 & $1.26\cdot 10^{-2}$ & 
$2.48\cdot 10^{-3}$ & 0 \\
${\rm BR}_{1,\,\rm exact}^{(V)}
$ & 0.25 & 0.104 &0.066 & 0.041 & $1.36\cdot 10^{-2}$ & 
$2.65\cdot 10^{-3}$ & 0 \\
${\rm BR}_{2,\,n=5}^{(A,V)}
$ &0.171 & 0.248 & 0.24 & 0.233 & 0.225 & 0.219 & 0.215 \\
${\rm BR}_{2,\, \rm exact}^{(A,V)}
$ &0.25 & 0.25 & 0.25 & 0.25 & 0.25 & 0.25 & 0.25 \\ 
$\gamma _{{\rm sl},\,n=5}/\gamma _{{\rm sl},\,\rm exact}
$  &.80 & 1.11 & 1.00 &0.97 & 0.94 
& 0.93 & 0.924 \\ 
\hline 
\end{tabular} 
\centering 
\caption{Comparing widths in the refined $n$ expansion with the 
exact results at $n=5$; the cases $m_c=0$ and $m_c\ge 0.2\,m_b$ are 
treated 
separately.}
\label{TABLE1}
\end{table} 
One sees that the $n$-expansion works with a typical accuracy 
of about $10\%$ for the inclusive width, as expected. The
weight of the nonleading terms decreases as 
$m_c \ra m_b$, i.e. in the SV limit. It is  remarkable
that the expansion based 
on the SV kinematics 
works so well down to a rather small mass ratio of 
$m_q/m_b \simeq 0.2$! This illustrates the observation 
made above that it is the parameter $(m_b - m_c)/nm_c$ 
that describes the proximity to the SV limit. For further analysis 
we note that at $m_c/m_b=0.3$ the `saddle points' for $n=5$ occur at
$\eta_*\equiv 1-\sqrt{q^2}/(m_b-m_c)$ equal to $0.16$, $0.28$ and 
$0.4$
for $\gamma_1^{(A)}$,  $\gamma_1^{(V)}$ and $\gamma_2$, 
respectively; for the $b\ra u$ case $\eta_{*}$ is $0.33$ and $0.55$ 
for $\gamma_{1,2}$, respectively.

Another cross-check of the numerical 
reliability of the refined $1/n$ expansion at $n=5$ is a comparison 
with the known perturbative expression at the 
one-loop level. In the SV regime one finds 
through order $\as/\pi $    
\beq
\gamma_{\rm sl,SV}\simeq 
\frac{8}{15}\Delta^5\;\left( 0.924-0.945\cdot \frac{\alpha_s}{\pi}
\right) = 
\frac{8}{15}\Delta^5\cdot 0.924
\left( 1-1.023\cdot \frac{\alpha_s}{\pi}\right) 
\label{13a}
\eeq 
to be compared to the exact result to that order  
\beq
\gamma_{\rm sl,exact}\simeq 
\frac{8}{15}\Delta^5\,\left( 1-\frac{\alpha_s}{\pi}\right)  
\;\; , 
\label{13b}
\eeq 
i.e., a difference of two percent only.

From these comparisons we conclude that the refined $1/n$ 
expansion yields 
good numerical results already for the physical value $n=5$.


\begin{thebibliography}{99}

\bibitem{svold}
M. Voloshin and M. Shifman, {\it Yad. Fiz.} {\bf 41} (1985) 187
[{\it Sov. J. Nucl. Phys.} {\bf 41} (1985) 120].

\bibitem{cgg}
J. Chay, H. Georgi and B. Grinstein, {\it Phys. Lett.} {\bf B247} (1990)
399.

\bibitem{mirage}
I. Bigi and N. Uraltsev, {\it Phys. Lett.} {\bf B 280} (1992) 271.

\bibitem{buv}
I. Bigi, N.G. Uraltsev and A. Vainshtein, {\it Phys. Lett.} {\bf B293}
(1992) 430; (E) B297 (1993) 477.

\bibitem{bs}
B. Blok and M. Shifman, {\it Nucl. Phys.} {\bf B399} (1993) 441 and 
459.

\bibitem{dpf}
I. Bigi, B. Blok, M. Shifman, N. Uraltsev and A. Vainshtein, {\it
The Fermilab Meeting}, Proc. of
the 1992 DPF meeting of APS, C.H. Albright {\it et al.} (World
Scientific, Singapore 1993), vol. 1, p. 610.

\bibitem{prl}
I. Bigi, M. Shifman, N. Uraltsev and A. Vainshtein, {\it Phys. Rev.
Lett.} {\bf 71} (1993) 496.

\bibitem{pole}
I. Bigi, M. Shifman, N. Uraltsev and A. Vainshtein,
{\it Phys. Rev.}  {\bf D50} (1994) 2234.

\bibitem{bbz}
M. Beneke, V. Braun and  V. I. Zakharov, {\it Phys. Rev. Lett.} {\bf 
73} (1994) 
3058. 

\bibitem{bbbsl}
P. Ball, M. Beneke and V.M. Braun,
{\it Phys. Rev.} {\bf D52 } (1995) 3929.

\bibitem{witten}
E. Witten, {\it The 1/N Expansion in Atomic and Particle Physics},
in Proc. Cargese Summer Institute ``Recent Developments in Gauge
Theories", Ed.  G. 't Hooft et al.  N.Y., Plenum Press, 1980.
 
\bibitem{SV}
M. Voloshin and M. Shifman, {\it Yad. Fiz.} {\bf 47} (1988) 801
[{\it Sov. J. Nucl. Phys.} {\bf 47} (1988) 511].

\bibitem{KOYRAKH}
B. Blok, L. Koyrakh, M. Shifman and A. Vainshtein,
{\it Phys. Rev. } {\bf D49} (1994) 3356.

\bibitem{optical}
I. Bigi, M. Shifman, N. Uraltsev and A. Vainshtein, 
{\it Phys. Rev.} {\bf D52} (1995) 196.

\bibitem{Volopt}
M. Voloshin, 
{\it Phys. Rev.} {\bf D46} (1992) 3062.

\bibitem{vcb}
M. Shifman, N. Uraltsev and A. Vainshtein, {\it Phys. Rev.} {\bf D51} 
(1995)
2217. 

\bibitem{motion}
I. Bigi, M. Shifman, N. Uraltsev and A. Vainshtein, 
{\it Int. Journ. Mod. Phys.} {\bf A9} (1994) 2467.

\bibitem{look}
M. Shifman and N.G. Uraltsev, {\it Int. Journ. Mod. Phys.} {\bf A10} 
(1995)
4705.

\bibitem{upset}
N.G. Uraltsev, {\em Int. Journ. Mod. Phys.} {\bf A11} (1996) 515.
\bibitem{bn}
P.~Ball and U.~Nierste, {\it Phys.Rev.} {\bf D50} (1994) 5841.

\bibitem{IW}
N. Isgur and M. Wise, {\it Phys. Lett.,} {\bf B232} (1989) 113;
{\it Phys. Lett.,} {\bf B237} (1990) 527.

\bibitem{dual}
B. Chibisov, R. Dikeman, M. Shifman and N.G.~Uraltsev, 
preprint  CERN-TH/96-113 [hep-ph/9605465]; to appear in {\em Int. 
Journ. Mod.
Phys. A} (1996).

\bibitem{pert}
N.G. Uraltsev, {\it Nucl. Phys.}, to appear [hep-ph/9610425].

\bibitem{rev}
I.~Bigi, M. Shifman and N. Uraltsev, preprint TPI-MINN-97/02-T 
[hep-ph/9703290].

\bibitem{wise}
M. Luke, M. Savage and M. Wise, {\it Phys. Lett.} {\bf B343} (1995) 329;
{\bf B345} (1995) 301.

\bibitem{BLM}
S.J. Brodsky, G.P. Lepage and  P.B. Mackenzie, {\it Phys. Rev.} {\bf
D28} (1983) 228;
for more references to the original approach and the extended 
summation see, e.g., \cite{bbbsl}.

\bibitem{czar}
A. Czarnecki, {\it Phys. Rev. Lett.} {\bf 76} (1996) 4124; \\
A. Czarnecki and K. Melnikov, hep-ph/9703277.

\bibitem{czarmel}
A. Czarnecki and K. Melnikov, Preprint Karlsruhe TTP 97-05, hep-
ph/9703291.

\end{thebibliography}
\end{document}